\documentstyle[12pt,lathuile]{article}
\newcommand{\be}{\begin{eqnarray}}
\newcommand{\ee}{\end{eqnarray}}
\newcommand{\bi}{\bibitem}
\newcommand{\nue}{\nu_e}
\newcommand{\num}{\nu_\mu}
\newcommand{\nut}{\nu_\tau}
\newcommand{\nus}{\nu_s}

\newcommand{\rar}{\rightarrow}
\newcommand{\lrar}{\leftrightarrow}

\newcommand{\tv}{\theta_{vac}}
\newcommand{\ds}{\partial \!  \! \! /}
\newcommand{\dm}{\delta m^2}
\newcommand{\raa}{\rho_{aa}}
\newcommand{\rss}{\rho_{ss}}
\newcommand{\stm}{\sin 2\theta_{m}}
\begin{document}
\title{ 
NEUTRINOS, OSCILLATIONS, AND BIG BANG NUCLEOSYNTHESIS
}
\author{
A.D. Dolgov        \\
{\em INFN, Ferrara, Italy and ITEP, Moscow, Russia } \\
}
\maketitle
\baselineskip=14.5pt
\begin{abstract}
The role of neutrinos in big bang nucleosynthesis is reviewed. Neutrino
oscillations in the early universe both in resonance and non-resonance 
case are briefly discussed. BBN and supernova limits on heavy sterile
neutrinos with the mass in 10-100 MeV range are presented and compared with
direct experimental bounds.
\end{abstract}
\baselineskip=17pt
\newpage
\section{Basics of BBN }
Big Bang Nucleosynthesis is known as one of the three solid pillars of 
the Standard Cosmological Model (SCM); the other two are the Cosmic
Microwave Background Radiation (CMBR) and General Relativity (GR). BBN
describes creation of light elements in the early hot universe in a very
(or rather, depending upon the attitude) good agreement with astronomical
observations. The theory predicts mass fraction  of $^4 He$ (with respect 
to total baryonic matter) about 25\%, the relative number densities of
deuterium, $^2 H$, and $^3 He$ at the level a few$\times 10^{-5}$ each, and
a few$\times 10^{-10}$ of $^7 Li$. The predictions span 9 orders of 
magnitude and well fit the data.

The characteristic temperature and time scales for the processes of light
element formations are respectively $T = (\sim{\rm MeV}) - 70$ keV and
$t = 0.1 - 10^3$ sec. The relation between the time and temperature (the
cooling rate) at this stage is given by
\be
(t/{\rm sec})(T/{\rm MeV})^2 = 
0.74 \left({10.75 \over g_*} \right)^{1/2}
\label{tT2}
\ee
Where $g_*$ describes the particle content of the primeval plasma. In the
SCM it is equal to 10.75 with the following contributions: 2 from photons, 
7/2 from electron-positron pairs, and $3\cdot 7/4$ from three neutrino
flavors. Any other form of energy present in the plasma is parametrized
by the contribution into $g_*$ as additional neutrino flavors,
$\delta g_* =(7/4)(N_\nu -3)$, though these hypothetical forms of energy
could be either in the form of massive particles, or vacuum-like energy,
or anything else. Because the cosmological cooling rate depends upon
$g_*$, BBN is sensitive to {\it any} form of energy from MeV down to
tens of keV range. 

There are of course some baryons in the plasma, building material for light
nuclei, but they do not have a noticeable contribution into the total energy
density because they are quite rare. Their number density, $n_B$,
is characterized by the parameter
\be
\eta_{10} = {n_B \over n_\gamma} = {\rm a\,\,\, few}\times 10^{-10} 
\label{eta10}
\ee
As we see in what follows, the output of light elements produced at BBN
depends upon $\eta_{10}$. It is the only unknown parameter of the 
standard theory. In fact, the data on light element abundances
are used to determine $\eta_{10}$. Till the last year it was the
only reasonably accurate way. Direct astronomical observations of the 
fraction of baryonic matter in the universe give about 10\% of the 
necessary amount of baryons. A dominant part of baryons is not visible. 
This year the measurements of the relative heights of the first and second
peaks in the angular fluctuations of CMBR permitted to determine $\eta_{10}$
independently\cite{2ndpeak}:
\be
\eta_{10}^{CMBR} =5.7 \pm 1.0 
\label{etacmbr}
\ee
This value is rather close to $\eta_{10}$ determined from 
BBN (see below).

\begin{figure}[ht]
\vspace{9.0cm}
\includegraphics{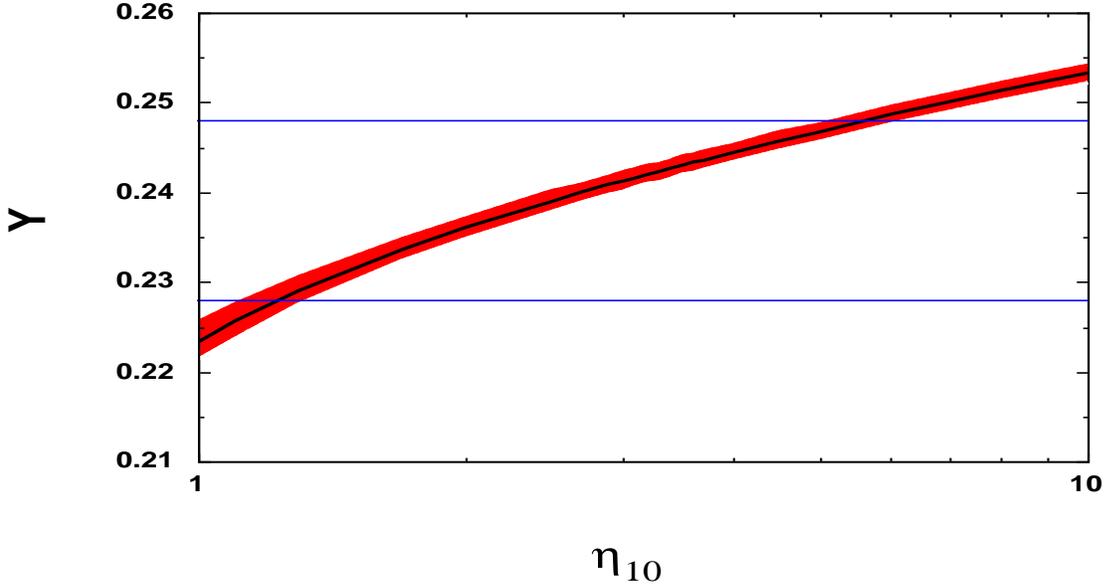}
	\caption{\it The predicted $^4 He$ abundance (solid curve) and 
the $2 \sigma$ theoretical uncertainty\protect\cite{hata}. The horizontal 
lines show the range indicated by the observational data.
	\label{Y}}
\end{figure}

\begin{figure}[ht]
 \vspace{9.0cm}
\includegraphics{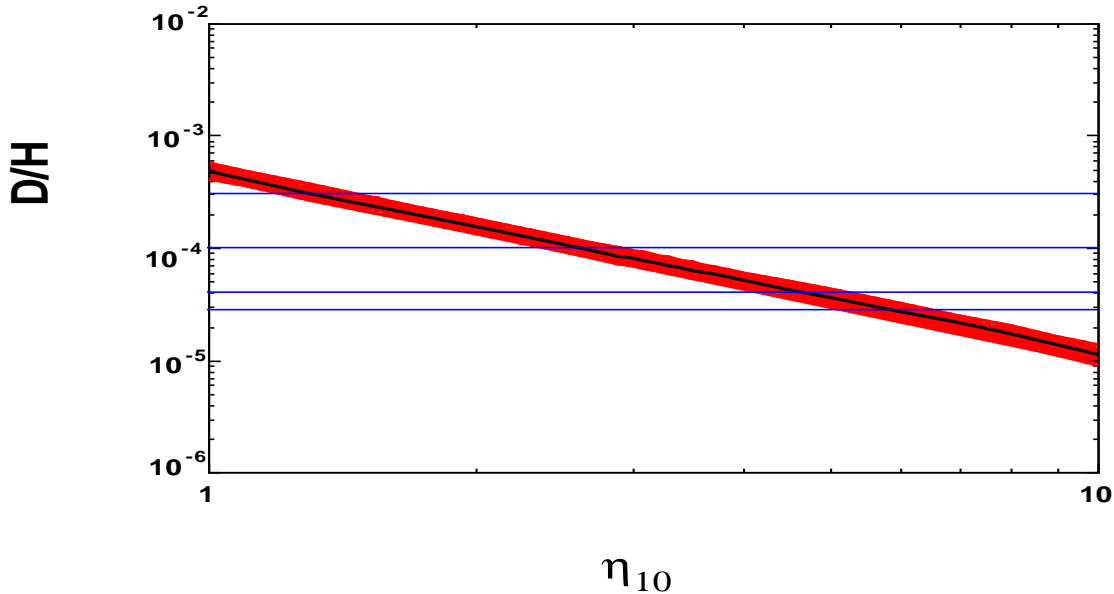}
	\caption{\it The predicted D/H abundance (solid curve) and the 
$2 \sigma$
theoretical uncertainty\protect\cite{hata}. The horizontal lines show the
range indicated by the observational data for both the high D/H (upper two
lines ) and low D/H (lower two lines).}
	\label{D}
\end{figure}

\begin{figure}[ht]
\vspace{9.0cm}
\includegraphics{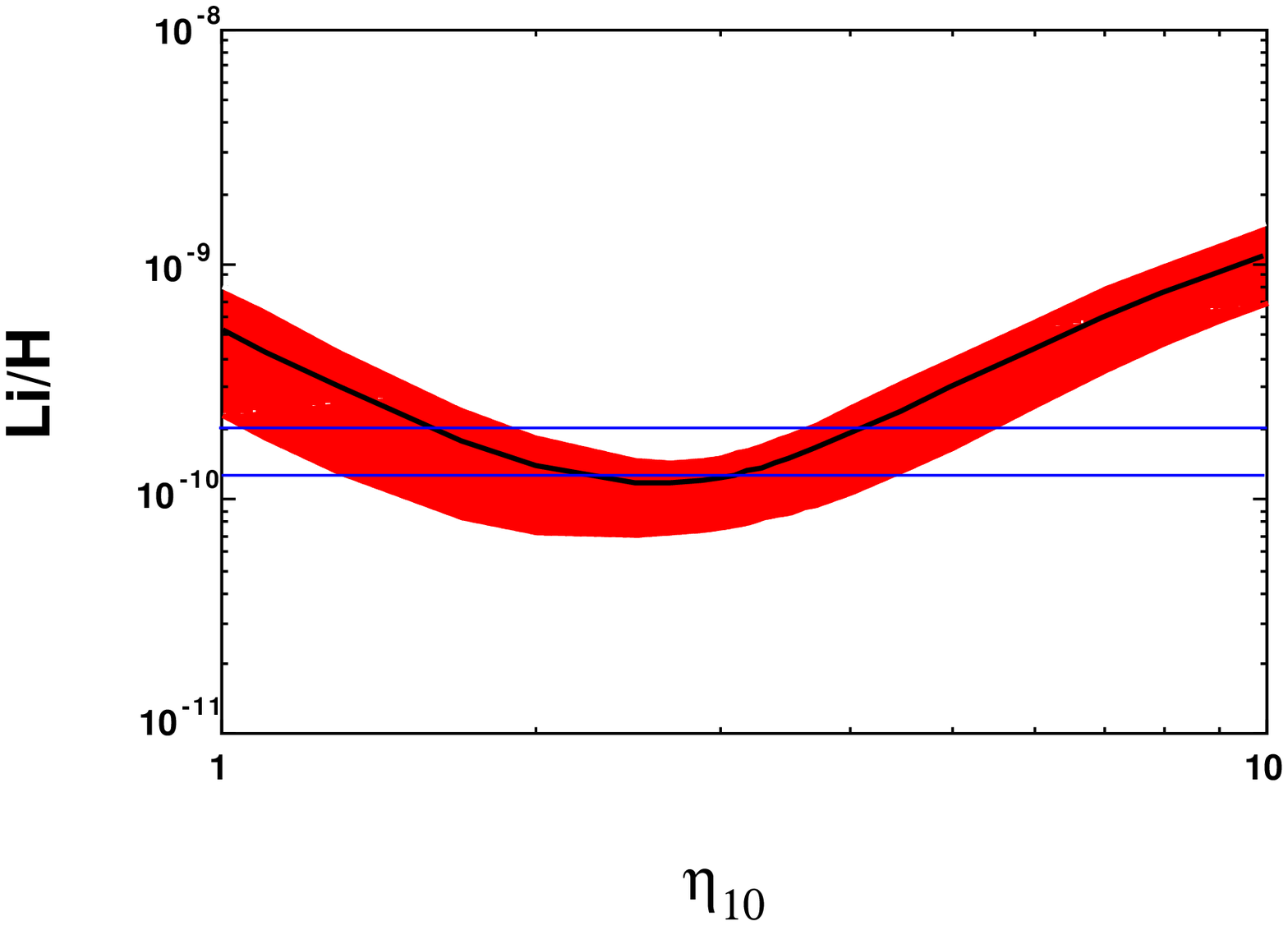}	
	\caption{\it The predicted $^7 Li$ abundance (solid curve) and 
the $2 \sigma$
theoretical uncertainty\protect\cite{hata}. The horizontal lines show the
range indicated by the observational data.}
	\label{Li}
\end{figure}

Production of light elements crucially depends upon the available number
of neutrons. The latter is determined by the competition between the 
rate of the reactions of neutron-proton transformation
\be
n+\nu_e \lrar p + e^-,\,\,\,
n+e^+ \lrar p +  \bar \nu
\label{nptran}
\ee
and the cosmological expansion rate,
\be
H = 5.44 \sqrt{{g_* \over 10.75}}\,\, {T^2 \over m_{Pl}}
\label{hubble}
\ee
where $m_{Pl} = 1.22\cdot 10^{19}$ GeV is the Planck mass. 
The rate of the reactions
is $\Gamma_{np} \sim G_F^2 T^5$ and at $T> 0.7$ MeV reactions are faster
than expansion and neutron-to-proton ratio follows equilibrium value,
$(n/p)_{eq} = \exp (-\Delta m/T)$, where 
$G_F = 1.166\times 10^{-5}$ GeV$^{-2}$ is the Fermi coupling constant 
and $\Delta m = 1.293$ MeV is the neutron-proton mass difference. When the
temperature drops below $\sim 0.7$ MeV the reaction become slow in 
comparison with expansion and the ratio $n/p$ would remain constant (frozen)
if not the neutron decay with the life-time $\tau_n=886.7$ sec. As a result
of the decay the neutron-to-proton ratio slowly decreases as
$\exp (-t/\tau_n)$. It goes on till the temperature drops down to the
temperature of nucleosynthesis, $T_{BBN} = 60-70$ keV, when formation 
of light elements abruptly and quickly begins. Practically all neutrons 
end their lives in $^4 He$ since the latter has the largest binding energy. 
Mass fraction of  $^4 He$ rather weakly, logarithmically, depends upon the 
number density of baryons, $\eta_{10}$. This quantity determines the 
temperature of BBN and respectively the number of neutrons that survived 
to this moment. One can see that $T_{BBN}$ is a very mild function of
$\eta_{10}$. On the other hand, helium-3 and especially deuterium are 
very sensitive to $\eta_{10}$ because these elements disappear in two-body 
baryon collisions whose probability is proportional to the number density 
of baryons. By this reason the amount of primordial deuterium serves as  
``baryometer'', measuring $\eta_{10}$. 
The calculated abundances of light elements as functions of $\eta_{10}$ 
are presented in figs.~\ref{Y}, \ref{D}, \ref{Li}.
taken from the review~\cite{olive99}. 

According to the recent measurements the observational data on primordial
mass fraction of $^4 He$ is concentrated near two centers
$Y_p = 0.234-0.238 $\cite{olive97,peimbert01,gruenwald01} and
$Y_p = 0.244$\cite{thuan00}. The first value corresponds to 
$\eta_{10} \approx 2$, while the other is close to 4-5. The analysis of
ref.~\cite{gruenwald01} was based on the data of the work~\cite{thuan00}
but with an account of ionization corrections.

Recent measurements of the number fraction of deuterium, $D/H$, in high 
red-shift Lyman-alpha clouds which presumably are not contaminated by
stellar processes and have its primordial quantity 
not diminished by burning is stars, give the results between 4 and 1.5 in
units of $10^{-5}
$\cite{omeara01,omeara01-2.53,dodorico01,pettini01,molaro99,levshakov01}.
The corresponding values of $\eta_{10}$ are between 4.5 and 8.5. 
There are also several measurements of very high deuterium values, 15-20
in the same units~\cite{webb97}, for the references to earlier papers
see e.g.~\cite{dolgov99}. However it was argued recently in 
ref.~\cite{kirkman01} that the system at red-shift $z=0.7$, where
the high deuterium value has been observed, has a complex velocity field
and cannot be used for identification of deuterium because the deuterium
line corresponds to velocity -81 km/sec and can be easily mimicked by 
the turbulent velocity field. It is concluded in the paper that high
deuterium is practically excluded.

\section{Role of neutrinos in BBN}

\begin{figure}[t]
\vspace{9.0cm}
\includegraphics{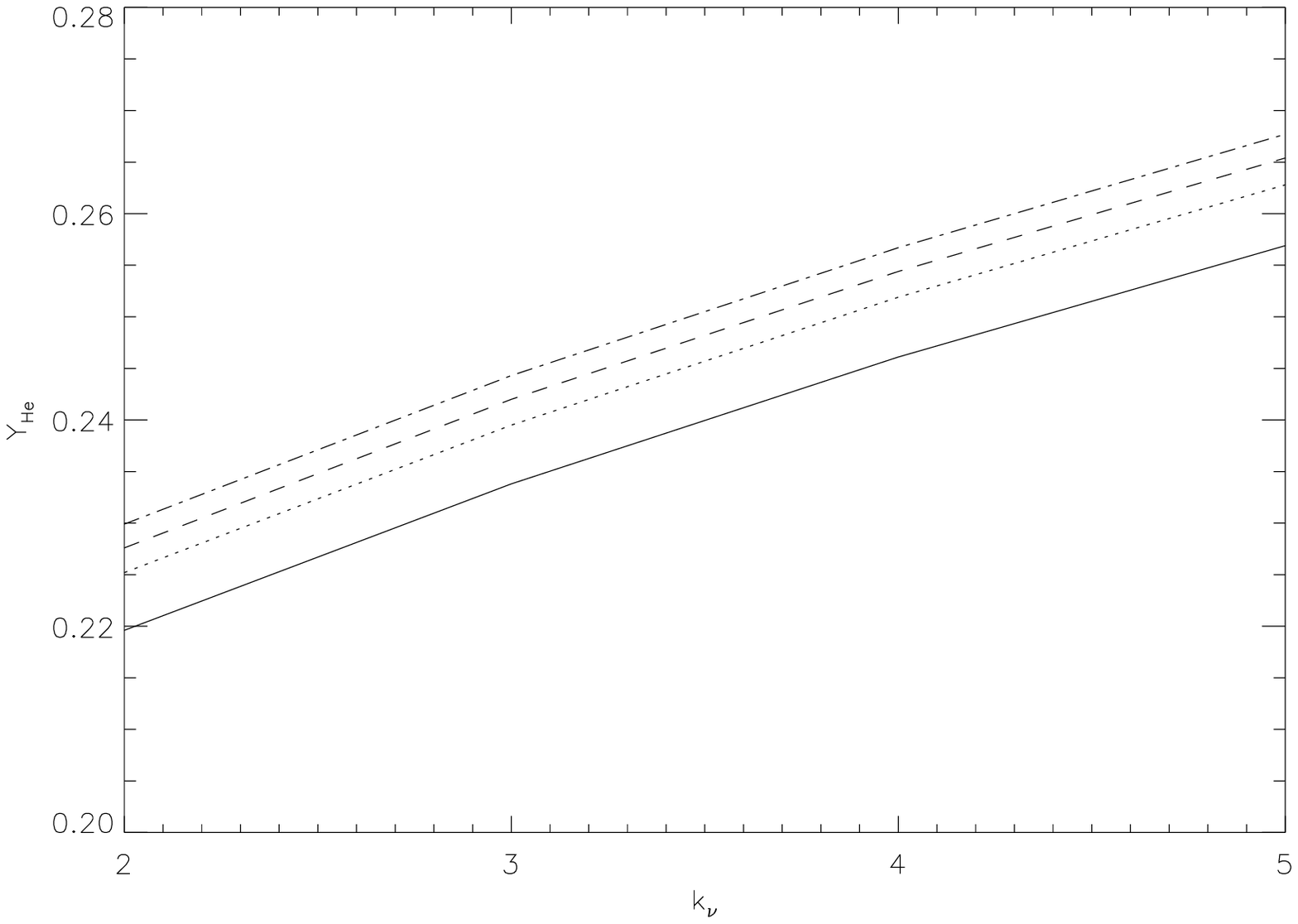}
 \caption{\it Mass fraction of $^4 He$ as a function of the number of massless
neutrino species. Different curves correspond to different values of the 
baryon-to-photon ratio $\eta_{10} \equiv 10^{10}n_B/n_\gamma =
2$  (solid),  3.1 (dotted), 4 (dashed),
and 5 (dashed-dotted).
    \label{helknu} }
\end{figure}

\begin{figure}[t]
 \vspace{9.0cm}
\includegraphics{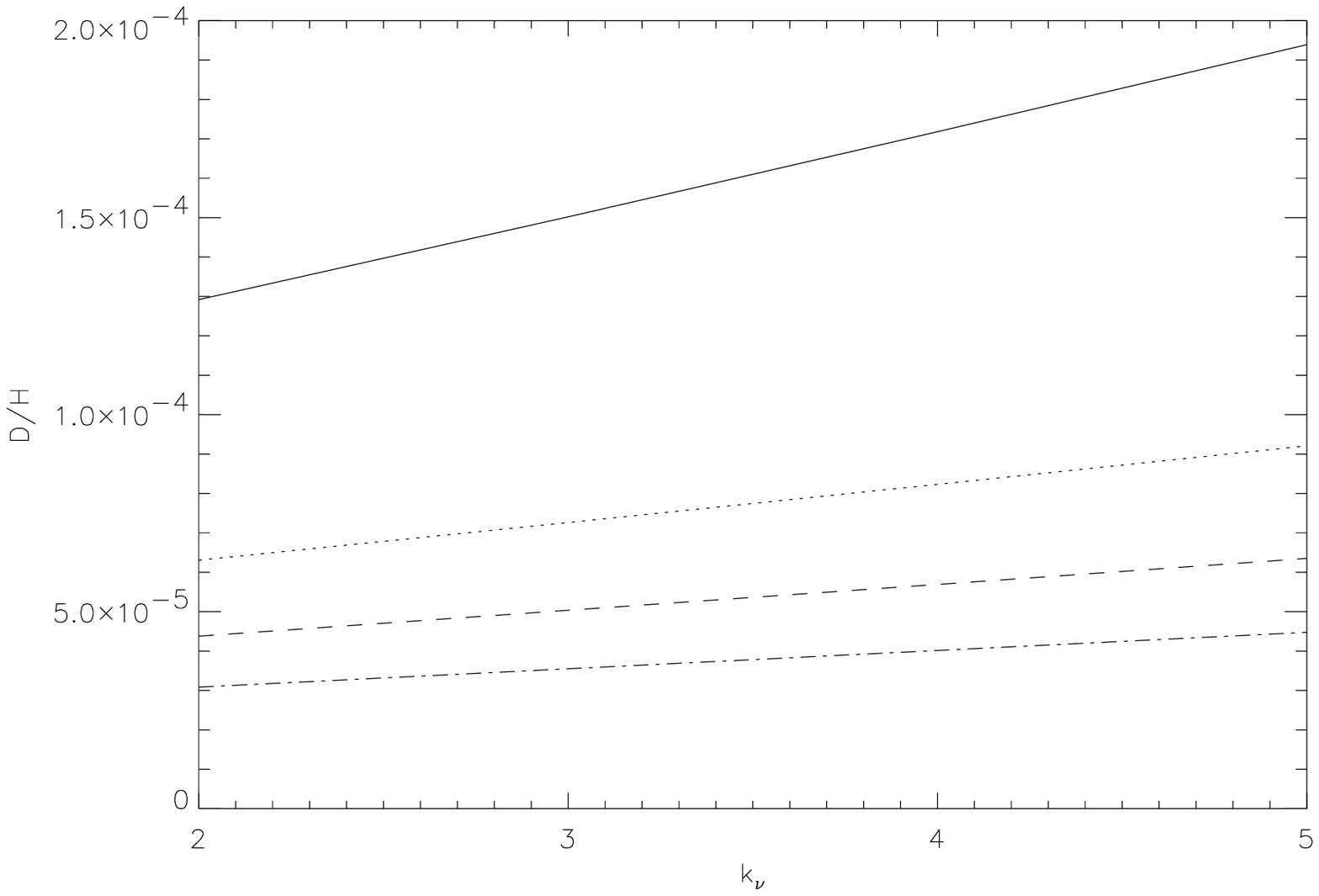}
 \caption{\it Deuterium-to-hydrogen by number as a function of the 
number of massless
neutrino species. Notations are the same as in fig. (\ref{helknu}).
    \label{deuterknu} }
\end{figure}

There are several physical effects that make neutrinos important for the
nucleosynthesis. First, as we have already mentioned above, neutrinos 
contributes to the total cosmological energy density and through it to 
the cooling rate, see eq.~(\ref{tT2}). The freezing temperature of 
$n/p$-transformation depends on $g_*$ as $T_{fr} \sim g_*^{1/6}$ and
the larger is $g_*$, the larger becomes $n/p$-ratio and more $^4 He$
is produced. Another effect is a decrease of the time interval which 
is necessary to reach the nucleosynthesis temperature 
$T_{BBN}$ (the latter does
not depend on $g_*$). Both effects work in the same direction and result 
in an increase of the mass fraction of $^4 He$. An addition of one
extra neutrino species leads to the increase of $Y_p$ by approximately
5\%. For deuterium the effect is of the same sign and stronger, see 
figs.~\ref{helknu}, \ref{deuterknu}.

The energy distributions of neutrinos are assumed to have the 
Fermi-Dirac equilibrium form:
\be
f_{\nu_j} = \left[ 1 + \exp \left( E/T - \xi_j \right)\right]
\label{fnuj}
\ee
where $E$ is the neutrino energy, and $\xi_j$ is dimensionless chemical
potential, $\xi_j =\mu_j /T$, and $j$ denotes neutrino flavour,
$j=e,\,\mu,\,\tau$. In the standard model is assumed that lepton asymmetry 
of neutrinos is negligibly small or in other words $\xi_j=0$. The impact 
of chemical potentials on light elements abundances is two-fold. First, 
degenerate neutrinos, i.e. those with  a non-zero $\xi$, have a larger 
energy density, corresponding to:
\be
g_* = 10.75\left[1 +0.3488 \sum_j \left(2\left({\xi_j \over \pi}\right)^2 
+\left({\xi_j \over \pi}\right)^4 \right) \right]
\label{gstarxi}
\ee
Thus a non-zero $\xi$ results in an increase of $^4 He$ and deuterium.

The second effect is operative only for electronic neutrinos because 
the latter are directly involved in the reactions~(\ref{nptran}). If
$\nue$ are degenerate, then the equilibrium $n/p$ ratio is shifted by
the factor $\exp ( -\xi_e )$ and the effect may have either sign 
depending on the sign of $\xi_e$. The most accurate recent 
bounds~\cite{hansen01} on the values of the chemical potentials based on
a combined analysis of the angular spectrum of CMBR and BBN with
the Deuterium fraction $D/H = (3.0\pm 0.4)\cdot 10^{-5}$, are
\be
-0.01 < \xi_e < 0.2,\,\,\, |\xi_{\mu,\tau} | < 2.6
\label{xibound}
\ee

The third possible phenomenon is a non-equilibrium distortion of
neutrino spectrum. By the reasons specified above it has an especially
strong effect on BBN for the case of electronic neutrinos. Normally
neutrinos are well in thermal equilibrium which is not distorted by
the cosmological expansion even if neutrino interactions is switched off 
at low temperatures, $T< (1-2)$ MeV. The latter is true only for massless
(or very light) particles. The spectrum may be distorted if there are
some new massive particles (or some neutrinos are heavy) which can
decay or annihilate into light neutrinos after their decoupling. The 
effect could be large but even in the standard model with the
usual massless neutrinos a significant spectrum distortion 
exists~\cite{dolgov92,dodelson92}.
Initially at large temperatures neutrinos were in strong thermal contact
with electron-positron plasma, but when the temperature dropped down the
contact became weaker and neutrino and electromagnetic components of the
plasma became almost independent. Later
$e^+e^-$ annihilation at $T\leq m_e$ heats 
up the electromagnetic component of the plasma and the temperatures of
neutrinos and electrons become different. Residual annihilation of hotter
electrons, $e^+e^- \rar \bar \nu \nu$, distorts neutrino spectrum. The
effect was accurately calculated by the numerical solution of the
integro-differential kinetic equations.
The most precise calculations performed by two different groups and by 
different methods~\cite{dolgov97,esposito00} show a very good agreement.
According to their results an excess of the energy densities for $\nue$
and $\num$ and $\nut$ are respectively
\be
\delta \rho_{\nue} / \rho_{\nue} = 0.94\%\,\,\,
{\rm and}\,\,\, \delta \rho_{\num} / \rho_{\num} = 0.40\%
\label{deltarhonu}
\ee 
However the impact of this effect on $^4He$ is extremely small, at the
level of $10^{-4}$. On the other hand, such correction may be in
principle observed in high precision measurements of CMB anisotropies 
by future PLANCK satellite mission~\cite{gnedin98,lopez98}, if the 
canonical model can be tested with the accuracy of about 1\% or better.
A change in neutrino energy density with respect 
to the standard case would result in a shift of equilibrium epoch between
matter and radiation, which is imprinted on the form of the angular spectrum
of fluctuations of CMBR. The total energy 
density of relativistic matter in the standard model is given by
\be
\Omega_{rel} = \Omega_\gamma \left[ 1 + 0.68\, {N_\nu \over 3}\,
\left( {1.401\, T_\nu \over T_\gamma}\right)^4 \right]
\label{omegarel}
\ee
where $\Omega_\gamma$ is the relative energy density of cosmic 
electromagnetic background radiation (CMBR) and $T_\gamma$ is the photon
temperature. The corrections discussed 
in this section and electromagnetic corrections of ref.~\cite{lopez98}
could be interpreted as a change of  $N_\nu$ from 3 to 3.04.

\section{Effects of neutrino oscillations on BBN}

In the case of oscillations between active neutrinos, i.e. $\nue$, $\num$,
and $\nut$, there would be no impact on BBN, if neutrinos were initially 
in thermal equilibrium with vanishing chemical potentials. Kinetic 
equilibrium is established automatically at $T>2 $ MeV due to a large
weak interaction rate in comparison with the expansion rate. Vanishing of
neutrino chemical potentials is the standard assumption but it is not
necessarily true. If neutrinos are degenerate, oscillations between 
different flavors could have an impact on the production of light elements.
The result depends not only on the parameters of the oscillations but also
on the initial values of the lepton asymmetries in different sectors. More
interesting physical effects originate in the case of oscillations between
active ($\nu_a$) and sterile ($\nu_s$) neutrinos. Initially $\nus$ should 
be absent in the primeval plasma. They can be produced only through the 
mixing with active ones but the mixing in matter is strongly suppressed
at high temperatures.

All three effects discussed in the previous section through which neutrinos
could influence BBN, are operative in the case of oscillations between
active and sterile neutrinos: 1) new relativistic species (namely $\nus$)
are created, enlarging $N_\nu$; 2) in the case of resonance oscillations
a large lepton asymmetry may be generated in the sector of active neutrinos;
3) spectrum of active neutrinos, especially of $\nue$, can be distorted by
oscillations.

If there exist one or several types of sterile neutrinos, i.e. neutrinos 
which do not any interaction with other particles and manifest themselves
only through non-diagonal mixing in the mass matrix, the eigenstates
of the free Hamiltonian can be parametrized in the following way:
\be
\nu_1 = \nu_a \cos \tv + \nus \sin \tv, \,\,\,
\nu_2 = -\nu_a \sin \tv + \nus \cos \tv
\label{nu1nu2}
\ee
where $\tv$ is the mixing angle in vacuum. The effects of medium are 
described by the refraction index (or, what is practically the same, by
effective potential). The latter can be calculated from the the Dirac 
equation averaged over thermal bath of cosmic plasma. The equation,
roughly speaking, has the form:
\be
i \ds \psi_a = {G_F \over \sqrt 2 }\, \left( \bar \psi_d 
O_\alpha \psi_b\right) O_\alpha \psi_a
\label{dpsia}
\ee
where $O_\alpha = \gamma_\alpha (1+\gamma_5)$. The factor in the r.h.s.
in front of $\psi_a$, averaged over plasma, gives the effective potential,
$(G_F/\sqrt 2) \langle \bar\psi O_\alpha \psi \rangle = 
\delta_{\alpha t} V_{eff}$.
The space component of this operator vanishes due to isotropy of the
plasma, while the time component is non-zero if the plasma is charge 
asymmetric. For the normally assumed magnitude of any cosmological charge
asymmetry at the level $10^{-9}-10^{-10}$ this term is subdominant and
the main contribution to refraction index comes from the so-called
non-local term. 

The operators $\psi_a$ in the l.h.s. and r.h.s. of
eq.~(\ref{dpsia}) may be at different space-time points because the weak
interaction is not a local one but is mediated by an exchange
of intermediate $W$ or $Z$ bosons. So the interaction is not the 
product of currents taken in the same space-time point.
The contribution of non-locality
is inversely proportional to $m_{W,Z}^2$ but still of the first order in
the coupling constant. Taking all contributions together we 
obtain~\cite{nora88}:
\be
V_{eff} = \pm C_1 \eta G_F T^3 + C_2^a G_F^2 T^4 E \alpha^{-1}
\label{Veff}
\ee
where $E$ is the neutrino energy, $\alpha = 1/137$, $C_1 = 0.95$, 
$C_2^e =0.61$, and   
$C_2^{\mu,\tau} = 0.61$. The numerical values of the coefficients $C_j$
are calculated for thermally equilibrium bath and for temperatures
below 100 MeV when muons were not present in the plasma. The coefficient 
$\eta$ is the charge asymmetry of the cosmological plasma,
including all particle species interacting with $\nu_a$:
\be
\eta^{(e)} = 
2\eta_{\nue} +\eta_{\num} + \eta_{\nut} +\eta_{e}-\eta_{n}/2 \,\,\,
 ( {\rm for} \,\, \nue)~,
\label{etanue} \\
\eta^{(\mu)} = 
2\eta_{\num} +\eta_{\nue} + \eta_{\nut} - \eta_{n}/2\,\,\,
({\rm for} \,\, \num)~,
\label{etanumu}
\ee
and $\eta^{(\tau)}$ for $\nut$ is obtained from eq.~(\ref{etanumu}) by 
the interchange $\mu \lrar \tau$. The individual charge asymmetries, 
$\eta_X$, are defined as the ratio of the difference between 
particle-antiparticle number densities to the number density of photons:
$ \eta_X = \left(n_X -n_{\bar X}\right) /n_\gamma $.
The charge asymmetry terms comes with negative sign for neutrinos and
with positive sign for antineutrinos.

At small $T$ the charge asymmetry term may be larger than the 
non-local one but at smaller temperatures the vacuum term, 
$\dm\cos 2\tv /2E$ dominates for $\dm > 10^{-7}$ eV$^2$. So for the 
``normal'' value of the cosmological charge asymmetry the first term
in eq.~(\ref{Veff}) is practically always subdominant.

An important difference between neutrino oscillations in stellar environment 
and in cosmology is that in the former the loss of coherence is not
important, neutrinos can be described by wave function and first
order effects (in terms of the coupling constant) given by the effective 
potential are sufficient for description of all essential physics. 
Situation in cosmology is more complicated. Neutrino production and 
annihilation, as well as 
elastic scattering at non-zero angle could be quite strong at high 
temperatures, so the coherence is quickly destroyed and the density matrix 
formalism should be used~\cite{dolgov81,stodolsky87,sigl93,mckellar94}. 
The kinetic equation for density matrix in the cosmological background 
has the form
\be
\dot\rho=\left( {\partial \over \partial t} - 
Hp {\partial \over \partial p} \right) \rho
&=& i\left[ {\cal H}_m + V_{eff}, \rho \right] +
\int d\tau (\bar \nu, l,\bar l) \left( f_l f_{\bar l} A A^+ -
{1\over 2} \left\{ \rho, A \bar\rho A^+ \right\} \right)+  \nonumber \\
&&\int d\tau (l,\nu',l') \left( f_{l'} B \rho' B^+ -
{1\over 2} f_l \left\{\rho, B B^+\right\} \right)
\label{decoh1}
\ee
where $\rho$ is the density matrix of the oscillating neutrinos,
$f_l$ is the distribution function of other leptons in the plasma,
$d\tau$ is the phase space element of all particles participating in
the reactions except for the neutrinos in question. It has essentially
the same form as the usual collision term in kinetic equation for 
the number density, when oscillations are absent:
\be
I^{coll}_i&=&-{(2\pi)^4 \over 2E_i}  \sum_{Z,Y}  \int  \,d\nu_Z  \,d\nu_Y 
\delta^4 (p_i +p_Y -p_Z) 
 \lbrack \mid A(i+Y\rightarrow  Z)  \mid^2  \nonumber \\
&&f_i  \prod_Y  f\prod_Z  (1\pm  f)-\mid  A(Z\rightarrow  i+Y)\mid^2 
\prod_Z f \prod_{i+Y} (1\pm f)\rbrack 
\label{si}
\ee
Here $Y$ and $Z$ are  arbitrary, generally  multi-particle  states, 
$\prod_Y f$ is  the  product  of  phase space densities  of  particles 
forming the state $Y$, and 
\be
d\nu_Y = \prod_Y {\overline {dp}} \equiv \prod_Y {d^3p\over (2\pi )^3 2E} 
\label{dnuy}
\ee
The signs '+' or '$-$' in $\prod (1\pm f)$ are chosen for  bosons  and 
fermions respectively.

The first commutator term in the r.h.s.
is first order in the interaction. It is the usual contribution from
refraction index which does not break coherence. The last two terms
are second order in the interaction and
are related respectively to annihilation, $\nu\bar\nu \lrar l\bar l$,
and elastic scattering, $\nu l \lrar \nu' l'$.
The quantum statistics factors, $(1-f)$, and $(I-\rho)$ are neglected
here. They can be easily reconstructed, see ref.~\cite{sigl93}.
In the interaction basis and in the case of active-sterile mixing
the matrices $A$ and $B$ have only one non-zero entry in the upper left
corner equal to the amplitude of annihilation or elastic scattering
respectively; the upper $'+'$ means Hermitian conjugate.

The exact form of the coherence breaking terms presented above is quite
complicated. In many cases an accurate description can be obtained with
a much simpler anzats:
\be
\dot \rho = ...- \{\Gamma, \left(\rho - \rho_{eq} \right)\}
\label{gammarho}
\ee
where the multi-dots denote contributions from the neutrino mixing in
medium described by the commutator term in eq.~(\ref{decoh1}),
$\rho_{eq}$ is the equilibrium value of the density matrix, i.e.
the unit matrix multiplied by the equilibrium distribution 
function~(\ref{fnuj}), and the matrix $\Gamma$ that describes the 
interaction with the medium, is diagonal in the flavor basis; it is 
expressed through the reaction rates. If we take for the latter the 
{\it total} scattering rate, including both elastic scattering and 
annihilation, we obtain in the Boltzmann approximation~\cite{harris82}:
\be
\Gamma_0 = 2\Gamma_1  = g_a \frac{180 \zeta(3)}{7 \pi ^4}
\, G_F^2 T^4 p  ~.
\label{gammaj1}
\ee
In general the coefficient $g_a(p)$ is a momentum-dependent
function, but in the approximation of neglecting $[1-f]$ factors in the
collision integral it becomes a constant~\cite{bell99} equal respectively
to $g_{\nu_e} \simeq 4$ and $g_{\nu_\mu,\mu_\tau} \simeq
2.9$ \cite{enqvist92b}. In ref.~\cite{dolgov00} more accurate values
are presented that are found from the thermal averaging of the complete
electro-weak rates (with factors $[1-f]$ included), which were 
calculated numerically from using the Standard Model code
of ref.~\cite{dolgov97}. This gives $g_{\nu_e} \simeq 3.56$
and $g_{\nu_\mu,\mu_\tau} \simeq 2.5$.

Let us first consider non-resonance oscillations. The effective mixing angle
in the medium is given by
\be
\sin 2\theta_m = { a\sin 2\tv \over a\cos 2\tv - V_{eff}}
\label{sin2tm}
\ee
where $a= \dm /2E$. In the pioneering papers~\cite{barbieri90,kainulainen90}
and in many subsequent ones the production rate of sterile neutrinos was 
estimated as:
\be
\Gamma_s = \langle  \Gamma_a \sin^2 2\theta_m \, \sin^2 {\dm \,t \over 2E}
 \rangle
\label{gammas}
\ee
where $\Gamma_a$ is the interaction rate of active neutrinos and
after averaging over time one can substitute 
$\langle \sin^2 ({\dm \,t / 2E}) \rangle = 1/2$. Demanding 
that the energy density
of $\nus$ produced by oscillations does not exceed the energy corresponding 
to $\Delta N_\nu$ neutrino species, one can obtain the following limit
\be
\dm\, \sin^4 \tv < 6\cdot 10^{-3} \left( T_a \over 3 \,{\rm MeV}\right)^6
\Delta N_\nu^2 \,\,{\rm eV}^2,
\label{dmsin4}
\ee
where $T_a$ is the decoupling temperature of active neutrinos, $\nu_a$.
In ref.~\cite{barbieri90} the latter was taken 3 MeV for $\nue$ 
and 5 MeV for
$\nu_{\mu,\tau}$. It corresponds to account only of the annihilation rate.
In ref.~\cite{kainulainen90} the decoupling temperatures were $T_e = 1.8$ eV
and $T_{\mu,\tau} = 2.7$ eV, corresponding to account of the total reaction
rate. To resolve the problem one has to turn to the kinetic equations for
the density matrix with the exact form of the coherence breaking 
terms~(\ref{decoh1}). Introducing real and imaginary parts for the
non-diagonal elements of the density matrix according to
$\rho_{as} =\rho^*_{sa} = R+iI$, one finds:
\be
\dot \raa (p_1) &=& -F I - 
\int A^2_{el} \left[ \raa (p_1) f_l(p_2) - \raa (p_3) f_l(p_4)\right]-
\nonumber \\
&&\int A^2_{ann} \left[\raa(p_1) \bar\raa (p_2)-f_l(p_3)f_{\bar l}(p_4)
\right],
\label{draa} \\
\dot \rss (p_1) &=& F I,
\label{drss} \\
\dot R (p_1) &=& WI - (1/2)\, R(p_1) \left[
\int d\tau (l_2,\nu_3,l_4) A^2_{el} f_l(p_2)\, + \right.
\nonumber \\
&&\int d\tau(\bar\nu_2,l_3,\bar l_4) A^2_{ann} \bar\raa (p_2) 
\left. \right],
\label{dR} \\
\dot I (p_1) &=& -WR - (F/2) \left(\rss -\raa \right)
- (1/2)\, I(p_1) \left[
\int d\tau (l_2,\nu_3,l_4) A^2_{el} f_l(p_2)\, + \right.
\nonumber \\
&&\int d\tau(\bar\nu_2,l_3,\bar l_4) A^2_{ann} \bar\raa (p_2) 
\left. \right],
\label{dI} 
\ee
The integration is taken over the phase space according to:
\be
d\tau = {1\over 2E_1} \int {d^3 p_2 \over (2\pi)^3 2E_2}
{d^3 p_2 \over (2\pi)^3 2E_2} {d^3 p_2 \over (2\pi)^3 2E_2}
(2\pi)^4 \delta^4 \left( p_1+p_2-p_3-p_4\right)
\label{int}
\ee
The amplitude of elastic scattering and annihilation with proper 
symmetrization factors can be taken from tables of ref.~\cite{dolgov97}. 

We will formally solve 
equations~(\ref{dR},\ref{dI}) to express the real and 
imaginary parts $R$ and $I$ of the non-diagonal components
through the diagonal ones. The relevant equations can be written as
\be
\dot R &=& WI - \Gamma_0 R/2 \label{aprdR} \,  , \\
\dot I &=& - WR + \frac{F}{2} \left( \raa - \rss \right) - \Gamma_0 I/2 \, .
\label{aprdI} 
\ee
In the limit when the oscillation frequency 
\be
\omega_{osc} = \left( F^2 +W^2 \right)^{1/2}
\label{omegaosc}
\ee
is much larger than the expansion rate, the solution is given by 
stationary point approximation, i.e. by the condition of vanishing
the r.h.s.:
\be
R = { FW \over 2\left( W^2 + \Gamma^2_0/4\right)} \left(
\rho_{aa} -\rho_{ss} \right)
\label{rstp}\\
I = { F\Gamma_0 \over 4 \left( W^2 + \Gamma^2_0/4\right)} \left(
\rho_{aa} -\rho_{ss} \right)
\label{istp}
\ee
In the non-resonant case, when $W\neq 0$, usually the condition
$W^2 \gg \Gamma^2/4$ is fulfilled and
\be
R \approx (\sin 2\theta_m /2)\left(\rho_{aa} -\rho_{ss} \right)
\label{rappr}\\
I \approx ({\sin 2\theta_m \Gamma_0 / 4W}) 
\left( \rho_{aa} -\rho_{ss} \right) 
\label{iappr}
\ee
where $\theta_m$ is the mixing angle in matter and in the limit of
small mixing
\be
\tan 2\theta_m \approx \sin 2\theta_m \approx {F\over W}
= {\sin 2\theta \over \cos 2\theta + \left( 2EV_{eff}/\dm \right)}
\label{sinthetam}
\ee
Now we can substitute the expression for $I$ into 
eqs.~(\ref{draa},\ref{drss}) and obtain a closed system of 
equations for the two unknown diagonal elements of density matrix which 
is easy to integrate numerically. In the case that the number density 
of sterile neutrinos is small and the active neutrinos are close to
equilibrium, as often the case, we obtain the following equation
that describes the production of sterile neutrinos by the oscillations
\be
\dot \rho_{ss} \approx ({\sin^2 2\theta_m \Gamma_0 / 4}) f_{eq}
\label{sprod}
\ee

An important conclusion of this derivation is that this production rate 
is by factor 2 smaller than the approximate estimates used in 
practically all earlier papers. 
An explanation of this extra factor $1/2$ is
that the time derivative of $\rho_{ss}$ is proportional to imaginary part 
of the non-diagonal component of the density matrix and the latter is
proportional to $\Gamma_1 =\Gamma_0/2$. 

The equations for the diagonal elements of density matrix can be solved
analytically in the limit of small number density of sterile neutrinos,
$\rss \ll\raa$ and for the case of kinetic equilibrium of active neutrinos,
so that in Boltzmann approximation their distribution takes the form:
\be
\rho_{aa} = C(x) \exp (-y)
\label{rhoaaeq}
\ee
where dimensionless variables $x=1\,\, {\rm MeV} a(t)$
and $y = p a(t)$ are introduced, with $a(t)$ being the cosmological
scale factor, normalized at high $T$ as $a=1/T$. 
Here $C(x)$ can be understood as an effective chemical potential,
$C(x) = \exp [\xi(x)]$, same for $\nu$ and $\bar\nu$. A justification
for this approximation is a much larger rate of elastic scattering
(that maintains the form~(\ref{rhoaaeq}) of $\rho_{aa}$) with respect
to annihilation rate that forces $\xi$ down to zero (or $\xi =-\bar\xi$
in the case of non-zero lepton asymmetry). This approach is similar to 
the standard calculations of cosmological freezing of species.
Now we can integrate both both sides of eq.~(\ref{draa}) over $d^3y$ 
so that the contribution of elastic scattering disappears and the 
following ordinary differential equation describing evolution of $C(x)$
is obtained:
\be
{dC\over dx} =-{k_l\over x^4} \left[ C^2 -1 + 
C\,{10 (1+g_L^2+g_R^2)\over 24 (1+2g_L^2 +2g_R^2) } 
\int dy\,y^3 e^{-y} \sin^2 2\theta_m
\right] 
\label{dcdx}
\ee
The first term in the r.h.s. of this equation came from annihilation
and the second one from the oscillations; the contribution of elastic
scattering disappeared after integration over $d^3y_1$.

We have assumed above that $\stm \ll 1$ and thus the term 
$\sim (\stm)^2 dC/dx$ has been
neglected. It is a good approximation even for not very weak mixing.
The constants $k_l$ are given by
\be
k_l={8G_F^2 \left( 1+2g_L^2 +2g_R^2\right) \over \pi^3 H\,x^2 }
\label{K}
\ee
so that $k_e = 0.17$ and $k_{\mu,\tau} = 0.098$. The charge asymmetry term
in neutrino refraction index was also neglected. 
It is reasonable in the non-resonant case if the asymmetry 
has a normal value around $10^{-9}-10^{-10}$.

Eq. (\ref{dcdx}) can be solved analytically if $|\delta|=|1-C| \ll 1$
and after some algebra we obtain the following result for 
the increase of the effective number of neutrino
species induced by mixing of active neutrinos with sterile ones:
\be
\Delta N_\nu = {1
\over 9\pi^2}\,
{\sin^2 2\theta_{vac} \over \sqrt {\beta_l}}{G_F^2 (1+g_L^2+g_R^2) \over 
Hx^2} \,\, {g_* (T^{\nus}_{prod}) \over 10.75}
\label{dlnnunus}
\ee 
where 
\be
\beta_e = {2.34\cdot 10^{-8} \over \dm \cos 2\theta}\,\,\,
{\rm and}\,\,\, 
\beta_{\mu,\tau} = {0.65\cdot 10^{-8} \over \dm \cos 2\theta},
\label{beta}
\ee

Substituting numerical values of the parameters we obtain:
\be
(\dm_{\nue\nus}/{\rm eV}^2) \sin^4 2\theta_{vac}^{\nue\nus} = 
3.16\cdot 10^{-5} (g_*(T^{\nus}_{prod})/10.75)^3 (\Delta N_\nu)^2
\label{dmess2}\\
(\dm_{\num\nus}/{\rm eV}^2) \sin^4 2\theta_{vac}^{\num\nus} = 
1.74\cdot 10^{-5} (g_*(T^{\nus}_{prod})/10.75)^3 (\Delta N_\nu)^2
\label{dmmuss2}
\ee
Here another factor $g_*$ came from the Hubble parameter.

These results can be compared with other calculations. They
are approximately 2 orders of magnitude stronger than
those presented in refs.~\cite{barbieri90}, where too high
freezing temperature for weak interaction rates was assumed and the limit
was obtained: $ \dm \sin^4 2\theta <6\cdot 10^{-3} \Delta N_\nu^2$.
In ref.~\cite{kainulainen90} the limit was 
$ \dm \sin^4 2\theta <3.6\cdot 10^{-4} \Delta N_\nu^2$. (All these
are given for mixing with $\nue$.)
On the other hand, the limits obtained in ref.~\cite{enqvist92b}
are approximately 6 time stronger than those found
above~(\ref{dmess2},\ref{dmmuss2}). They are: 
$ \dm \sin^4 2\theta <5\cdot 10^{-6} \Delta N_\nu^2$ for
mixing with $\nue$ and 
$ \dm \sin^4 2\theta <3\cdot 10^{-6} \Delta N_\nu^2$ for mixing
with $\nu_{\mu,\tau}$. The difference by factor 6
between these results and 
eq.~(\ref{dmess2},\ref{dmmuss2}) can be understood in part by the factor
2 difference in the interaction rate, according to eq. (\ref{sprod}),
which gives factor 4 difference in the limits. The remaining difference 
by roughly factor 1.5 could possibly be prescribed to different ways of
solution of kinetic equations or to the fact that the increase in the number
density of sterile neutrinos is accompanied by the equal decrease in 
the number density of active neutrinos if the production of the latter by 
inverse annihilation is not efficient. This phenomenon is missed
in kinetic equations with the simplified form of the coherence
breaking term~(\ref{gammarho}), which are mostly 
used in the literature, while equations (\ref{draa}-\ref{dI})
automatically take that into account. However, for sufficiently large
mass difference, $\dm$, the effective temperature of $\nus$ 
production is:
\be
T^{\nus}_{prod} = (12-15)\, (3/y)^{1/3}\, 
(\dm/{\rm eV}^2)^{1/6}\,\, {\rm MeV}
\label{tprodnus}
\ee
Here the first number is for mixing of $\nus$ with $\nue$ and
the second one is for mixing with $\num$ or $\nut$. It
is larger than the temperature of the
annihilation freezing, so the active neutrino states are quickly
re-populated and the mentioned above  affect could be significant only for
a small mass difference. Much weaker bounds obtained in 
ref.~\cite{dolgov00c} resulted from an error in the coherence breaking
terms in kinetic equations for the non-diagonal matrix elements of the
density matrix.

The result~(\ref{dmess2}) includes only a rise of the total cosmological
energy density due to production of sterile neutrinos. In the case of
mixing with $\nue$ the distortion of spectrum of the latter is also 
important for BBN. This issue is addressed in 
refs.~\cite{kirilova97,kirilova98a,kirilova99b}.

\begin{figure}[t]
 \vspace{9.0cm}
\includegraphics{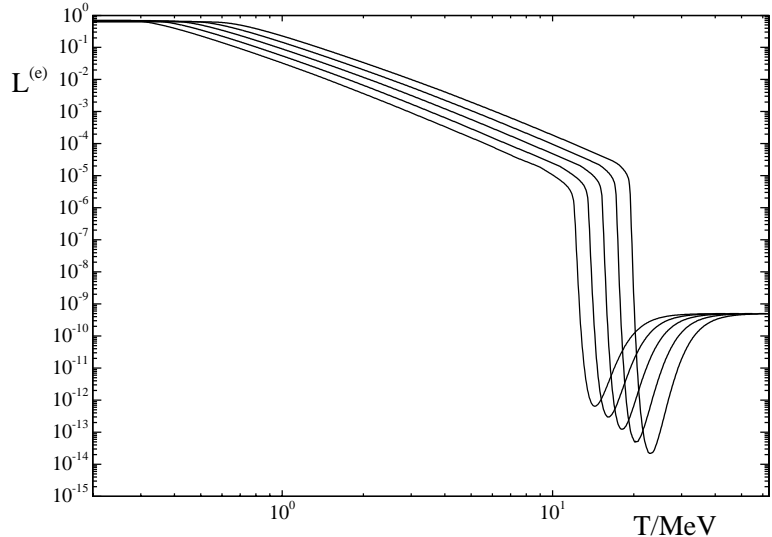}
 \caption{\it Evolution of $L^{(e)} = 2L_{\nu_e} + \eta$
for $\nu_e \to \nu_s$ oscillations with $\sin^2 2\theta_0
= 10^{-8}$ and, from left to right, $\delta m^2/eV^2 =
-0.25, -0.5, -1.0, -2.0, -4.0$ obtained
from numerically solution of the quantum kinetic
equations. The initial $L_{\nu_e} = 0$ is taken and 
$\eta = 5\times 10^{-10}$ is assumed.
The low temperature evolution is weakly dependent on these values.
\label{fig:dbfoot} }
\end{figure}

If the mass difference between $\nus$ and $\nu_a$ is negative then the 
MSW-resonance transition might take place in cosmological background. 
The resonance may also exist with an arbitrary sign of $\dm$ if the initial 
value of the asymmetry is sufficiently large but we will not consider 
this case. A very interesting phenomenon was discovered in 
ref.~\cite{ftv}. It was found that initially very small lepton asymmetry
($\sim 10^{-9}$) can rise almost to unity because of oscillations. The
effect is based on the dependence of the refraction index on the cosmological
charge asymmetry (the first term in eq.(\ref{Veff})). Since this term comes
with different signs to neutrinos and antineutrinos, the resonance
conditions for the same value of momenta could be reached e.g. earlier 
for $\nu$ than for $\bar\nu$. This would give rise to a more efficient
transformation of neutrinos in comparison with antineutrinos and the 
asymmetry could start rising. This creates a positive feed-back effect
and exponential rise of the asymmetry at initial stage. In fact, initially
the asymmetry very strongly drops down and only after some rather short
period the explosive rise begins. The first results were obtained by
numerical solution of kinetic equations for the density matrix, with 
some reasonable approximations made. In subsequent works more precise
numerical methods have been developed. The present-day state of art and
the list of relevant references can be found e.g. in the recent 
papers~\cite{foot99c,foot00,dibari01}. Numerical solution presents a
serious challenge because of quickly oscillating functions under the 
sign of the integral over neutrino momentum which enters the charge
asymmetry term. On the other hand, the presence of fast and slow variables
permits to separate them and to find an approximate analytical 
solution~\cite{dolgov01}. Numerical and analytical results are in 
a very good agreement in the coinciding range of parameters. 
The evolution of $\nue$-asymmetry according to calculations
of ref.~\cite{dibari00} for $\sin^2 2\theta = 10^{-8}$ and several
values of mass difference is presented in fig.~\ref{fig:dbfoot}.

Immediately after the large rise of lepton asymmetry due to oscillations
between active and sterile neutrinos was discovered, it was 
argued~\cite{shi96} that the resulting asymmetry is large but chaotic, 
i.e. its sign is essentially unpredictable.  The sign of the 
asymmetry is very sensitive to the oscillation parameters and the input 
of numerical calculations. As a result of this feature the sign
of lepton asymmetry might be different in different causally 
non-connected domains~\cite{shi99d1}. This could have interesting
implications and, in particular, would lead to 
inhomogeneous nucleosynthesis. The existence of chaoticity
was confirmed in several subsequent 
papers~\cite{enqvist99} but again in the 
frameworks of simplified thermally averaged equations. 
On the other hand, the analysis of possible chaoticity performed 
in ref.~\cite{dibari99c} on the basis of numerical solution of kinetic 
equations with a full momentum dependence shows a different picture. 
Most of parameter space is not chaotic, 
while in the region where chaoticity is observed numerical calculations
are not reliable. Analytical calculations of ref.~\cite{dolgov01} also
do not show any chaoticity. At the present time the problem remains 
unresolved.

A very interesting phenomenon of spatial fluctuations of lepton
asymmetry was found in ref.~\cite{dibari99b}.
Neutrino oscillations in the presence of initially small
baryonic inhomogeneities could give rise to domains
with different signs of lepton asymmetry. This effect is different
from chaotic amplification of asymmetry discussed above. As we have
already mentioned the initial asymmetry
first drops down to an exponentially small value and after that started 
to rise, also exponentially, with a larger integrated exponent.
Since the value of the asymmetry in the minimum could be extremely small,
it is sensitive to small perturbations and the final sign of asymmetry
could be determined by them. This goes not in a trivial way as e.g.
spatial fluctuations of the sign in the minimum but somewhat more tricky
and related to diffusion term in kinetic equations.
As is argued in the original paper~\cite{dibari99b} 
(see also ref.~\cite{enqvist00}), small spatial fluctuations of the
cosmological baryon asymmetry, though they do not create change of sign
in the minimum, would induce formation of domain with super-horizon
sizes (at the moment of their creation) with large lepton 
asymmetry of different signs.

Let us consider the equation, modeling evolution of the lepton
asymmetry in the presence of small spatial inhomogeneities, used in 
ref.~\cite{dibari99b}. Following notations of this paper let us denote
the asymmetry of active neutrinos of flavor $a$ as $L_{\nu_a}$. The
combination that enters the refraction index of $\nu_a$ is
$2 L_{\nu_a} + \tilde L$, where $\tilde L$ consists of the contributions
of other neutrino species, baryons, and electrons. Due to electric charge
neutrality the last two are not independent. The preexisting fluctuations 
in asymmetry of neutrinos would be erased due to a large neutrino mean free
path in cosmic plasma after neutrino decoupling. So the background
asymmetry can be written as $\tilde L = \bar L + \delta B (\vec x)$,
where the first term is homogeneous and the last one could be 
inhomogeneous and related to the fluctuations in the baryon number.
Now we will neglect the universe expansion (it is essential for quantitative
estimates but not for a qualitative result) and will solve the evolution
equations in somewhat different way, permitting more accurate evaluation
of the effects of diffusion.
The evolution of the neutrino asymmetry is described by the equation:
\be
\dot{L}_{\nu_{\alpha}}(\vec{x},t)=  
a(t) \left[ 2\,L_{\nu_{\alpha}}(\vec{x},t) + \bar L +
\delta B(\vec{x})\right ] 
+D(t)\nabla^{2}L_{\nu_{\alpha}}(\vec{x},t)
\label{dotLnu}
\ee
where $D(t)$ is the diffusion coefficient, and the function $a(t)$
is initially negative and generate an exponential decrease of the
asymmetry, but at some critical time $t_c$ it changes sign and this 
creates a huge rise of the asymmetry. It is essential that, while 
$a(t)$ is negative, the asymmetry drops down to a very small value.
In a more accurate formulation $a$ would also
depend on the asymmetry itself, but in what follows we are 
interested in rather small values of the asymmetry, where the non-linear 
effects are not important.

The solution to this equation can be found by the Fourier transform
and we obtain:
\be
L_{\nu_{\alpha}}(x,t) & = &\bar{L}\, \int_{t_{\rm in}}^t dt'\,a(t')
\,e^{2\int_{t'}^t\,dt''\,a(t'')} +\\ \nonumber
& +& e^{2\,\int_{t_{\rm in}}^t dt'\,a(t')}
\int\,d^3k\,e^{i{\vec k}{\vec x}}\,
\hat{L}_{\nu_a}({\vec k},t_{\rm in})\,e^{-k^2\,
\int_{t_{\rm in}}^t dt'\,D(t')}+ \\
\nonumber
& + & \int_{t_{\rm in}}^t dt'\,a(t')\,e^{2\int_{t'}^t\,
dt''\,a(t'')} \,\int \,d^3k\,e^{i{\vec k}{\vec x}}\,
\delta \hat{B}({\vec k},t_{\rm in})\,e^{-k^2\,
\int_{t'}^t dt''\,D(t'')}
\label{solltx}
\ee
Here ``hut'' indicates the Fourier transform of the corresponding function.
The first integral in this expression can be explicitly taken
because the integration measure $dt' a(t')$ is exactly the differential
of the exponential. The integration of this term gives:
\be
(1/2) \bar L \left[ \exp \left( 2\int^t_{t_{in}} dt_2 a(t_2) \right)
-1 \right]
\label{barL}
\ee
So we obtained a rising term (after some initial decrease) plus a constant
initial value of $\bar L$.

Moreover, the second term can be further integrated 
because the initial value $\hat{L}({\vec k},t_{\rm in})$ is
supposed to be homogeneous and so its Fourier transform is just
delta-function, $\delta^3 (k)$. The integral gives
\be
L_{\nu_a}^{(in)}\, \exp \left( 2\int^t_{t_{in}} dt_2 a(t_2) \right)
\label{Lnua}
\ee
where $L_{\nu_a}^{(in)} $ means the initial value, i.e.
taken at $t=t_{in}$. So if we forget about the constant term $\bar L/2$,
we would have $(L_{\nu_a}^{(in)} +\bar L/2)$
multiplied by the rising exponent. One would get exactly this expression 
if one solves the equation for $\dot L_{\nu_a}$ in homogeneous case.

Let us now consider the last term which, according to the 
arguments of ref.~\cite{dibari99b}, could change the
sign of the rising asymmetry, i.e. this last term could become the dominant 
one. To evaluate the integral let us substitute:
\be
\delta \hat{B}({\vec k},t_{\rm in}) = 
\int d^3 x_1 e^{i{\vec k}{\vec x}_1}\, \delta B ({\vec x}_1)
\label{deltabk}
\ee
where $\delta B ({\vec x}_1)$ is the initial value of the inhomogeneous 
term. Now we can make integration over $d^3 k$. We have the integral of 
the type
\be
\int d^3k \exp [- S^2 k^2 +i {\vec k} (\vec{x}-{\vec x}_1)]
\label{intdk}
\ee
the scalar product of vectors $\vec k$ and $\vec r = \vec x-\vec x_1$ 
is equal to ${\vec k}({\vec x}-{\vec x_1}) = kr\cos\theta$, 
and
\be
S^2 (t) = \int_{t_{in}}^t dt_2 D(t_2) 
\label{S}
\ee
Integration over angles in $d^3 k = 2\pi k^2 dk d(\cos\theta)$ is
trivial, it gives $\sin kr /kr$. The remaining integration can be done 
as follows:
\be
\int dk k \sin kr \exp [-S^2k^2] = 
(d/dr) \int dk \cos kr \exp [-S^2k^2]
\label{intk2}
\ee
and the last integral can be taken if we expand the range of integration 
from minus to plus infinity.
Introducing new variable 
$\vec x_1 = \vec x - S(t_1)\vec \rho$ we finally obtain
\be
\int dt_1 a(t_1)\, e^{2\int_{t_{1}}^t dt_2 a(t_2)} \int d^3 \rho \,
\delta B (\vec x -S(t_1) \vec \rho) \, e^{-\rho^2}
\label{fin}
\ee
This is the contribution the lepton asymmetry $L_{\nu_a}$ generated by 
the (small) baryonic inhomogeneities. Its asymptotic rise at large
$t$ is similar to the rise of other terms but its exponential decrease
at intermediate stage could be considerably milder and, as a result,
this term could become dominant with the sign determined by the
sign of the fluctuations in the baryon asymmetry. We can check this
on a simple example assuming that the function $a(t)$ has the form
$a(t) = a_1 (t-t_c)$ and that fluctuations of the asymmetry
are described by one harmonic mode: 
$\delta B({\vec x}) = \epsilon_B \cos {\vec k_0}{\vec x}$. This 
form of $\delta B$ could be inserted either into eq.~(\ref{fin})
or into initial eq.~(\ref{solltx}) and we find for the oscillating
part of the asymmetry (up to a constant coefficient):
\be
\delta L({\vec x}) &=& \epsilon_B \cos {\vec k_0}{\vec x} \,\,
e^{a_1 (t-t_c)^2 - S(t)\,k_0^2} 
\left[ \int_{t_c-t_{in}}^{t-t_c} dt_1 t_1
e^{-a_1 t_1^2 + S(t_1)\, k_0^2}+ \right.
\nonumber \\ 
&&\left.
\int_0^{t_c-t_{in}} dt_1 t_1 e^{-a_1 t_1^2} 
\left( e^{S(t_1)\, k_0^2 } - e^{-S(t_1)\, k_0^2 }\right)\right]
\label{deltaL}
\ee
Both terms rise as $\exp [a_1 (t-t_c)^2]$, i.e. in the same way as
the other homogeneous terms (we assume that $S(t)$ is finite at large
$t$ and not too large). The first term is exponentially suppressed
as $\exp [-a_1 (t_c-t_{in})]$ also at the same level as the homogeneous 
terms. The second term, which vanishes in homogeneous case ( $k_0 =0$
or $ S=0 $) is not exponentially suppressed. In the limit of large
$a_1$ the integral can be evaluated as $\sim S(0)\,k_0^2 /a_1^3$. It
is small but not exponentially small. Thus, it is easy to imagine the
situation when the last term dominates and resonance enhancement of
lepton asymmetry in the background of small fluctuations of baryon 
asymmetry could create domains with large and different lepton
asymmetry. The effect is very interesting and deserves more 
consideration.

\section{Heavy sterile neutrinos (with 10-100 MeV mass)}

A hypothesis that such neutrinos may exist originated from
the observation of the KARMEN anomaly  in the time distribution
of the charged and neutral current events induced by neutrinos from 
$\pi^+$ and $\mu^+$ decays at rest~\cite{karmen95}. A suggested 
explanation of this anomaly was the production of a new neutral particle 
in pion decay
\be
\pi^+ \rar \mu^+ + x^0~,
\label{piondecay}
\ee
with the mass 33.9 MeV, barely permitted by the phase space, so that this
particle moves with non-relativistic velocity. Its subsequent 
neutrino-producing decays could be the source of the delayed neutrinos 
observed in the experiment. Among possible candidates on the role of
$x^0$-particle was, in particular, a 33.9-MeV sterile 
neutrino~\cite{barger95}.  According to ref.~\cite{dolgov00krm},
cosmology and astrophysics practically exclude the 
interpretation of the KARMEN anomaly by a 33.9~MeV neutrino mixed 
with $\nut$. It agrees with a later statement by the KARMEN collaboration 
made at Neutrino 2000~\cite{karmen2000} that the anomaly 
was not observed in upgraded detector KARMEN 2, but the question
still remains which area in the mass-mixing-plane for heavy sterile
neutrinos can be excluded. This issue was addressed recently by NOMAD
collaboration in direct experiment~\cite{nomad00} and in 
ref.~\cite{dolgov00hv} by considerations of big bang nucleosynthesis and
the  the duration of the supernova (SN)~1987A neutrino burst.

The mixing of heavy sterile neutrinos with the active ones couples
the former to $Z^0$ boson and induce the decay
\be
\nu_2 \rightarrow \nu_1 + \ell + \bar \ell \, ,
\label{dech}
\ee
where $\ell$ is any lepton with the mass smaller than the mass $m_2$ 
of the heavy neutrino. If  $m_2 < 2 m_\mu$ the decay into $\bar\mu \mu$ 
and $\bar\tau \tau$ is kinematically forbidden. 
If $\nus$ is mixed either with  $\num$ or $\nut$, the life-time is 
expressed through the mixing angle as:
\be
\tau_{\nu_s}  \equiv \Gamma_{\nu_2}^{-1} = { 
1.0~{\rm sec} \over (M_s/\mbox{10 MeV})^5\, \mbox{sin}^2 2\theta}  \, .
\label{taumuh}
\ee
For the mixing with $\nue$ the numerator is 0.7~sec; the
difference is due to the charged-current interactions. 

There are several effects operating in different directions, by which
a heavy unstable sterile neutrino could influence big-bang 
nucleosynthesis. First, their contribution to the total energy density
would speed up the expansion and enlarge the frozen neutron-to-proton 
ratio. Less direct but stronger influence could be exerted through the
decay products, $\nue,\, \num$, and $\nut$, and $e^\pm$ and through
the change of the temperature evolution, $T_\nu/T_\gamma$. 
The impact of $\num$ and
$\nut$ on BBN is rather straightforward: their energy density increases
with respect to the standard case and this also results in an increase
of $r_n$. This effect can be described by the increased number of
effective neutrino species $N_\nu$ during BBN. 
The increase of the energy density of $\nue$, due to decay of 
$\nus$ into $\nue$, has an opposite effect on $r_n$. Though a larger energy
density results in faster cooling, the increased number of $\nue$
would preserve thermal equilibrium between neutrons and protons for a longer 
time and correspondingly the frozen $n/p$-ratio would become smaller.
The second effect is stronger, so the net result is a smaller $n/p$-ratio. 
There is, however, another effect of a distortion of the equilibrium
energy spectrum of $\nue$ due to $e^\pm$ produced 
from the decays of $\nus$. If the spectrum is distorted
at the high-energy tail, as is the case, then creation of protons in 
the reaction $n+\nue \rar p + e^-$ would be less efficient than neutron
creation in the reaction $ \bar \nue + p \rar n + e^+$. We found that
this effect is quite significant. Last but no the least, the decays of 
$\nus$ into the $e^+e^-$-channel will inject more energy into the 
electromagnetic part of the primeval plasma and this will diminish 
the relative contribution of the energy density of light neutrinos and
diminish $r_n$. 

In refs.~\cite{dolgov00krm,dolgov00hv} the distribution functions of 
neutrinos were calculated from kinetic equations in Boltzmann 
approximation and in a large part of parameter space they significantly
deviate from equilibrium. The distributions of electrons and positrons 
were assumed to be very close to equilibrium because of their very 
fast thermalization due to interaction with the photon bath. However, 
the evolution of the photon temperature, due to decay and annihilation of 
the massive $\nus$ was different from the standard one, 
$T_{\gamma} \sim 1/a(t)$, by an extra factor $(1+\Delta) >1$ and this 
was explicitly calculated from the energy balance condition. 
At sufficiently high temperatures, $T>T_W \sim 2$~MeV,
light neutrinos and electrons/positrons were in strong contact, so that
the neutrino distributions were also very close to the equilibrium ones. If
$\nus$ disappeared sufficiently early, while 
thermal equilibrium between $e^{\pm}$ and neutrinos remained, then 
$\nus$ would not have any observable effect on primordial abundances, 
because only the contribution of neutrino energy density relative to 
the energy density of $e^{\pm}$ and $\gamma$ is
essential for nucleosynthesis. Hence a very short-lived $\nus$ has a negligible
impact on primordial abundances, while with an increasing lifetime the
effect becomes stronger. Indeed at $T<T_W$ the exchange of energy between 
neutrinos and electrons becomes very weak and the energy injected into the 
neutrino component is not immediately redistributed between all the particles.
The branching ratio of the decay of $\nus$ into $e^+e^-$ is approximately
1/9, so that the neutrino component is heated much more than the 
electromagnetic one. As we mentioned above, this leads to a faster cooling
and to a larger $n/p$-ratio. 

If the equilibrium number density of sterile neutrinos is reached, it
would be maintained until $T_f \approx 4 (\sin 2\theta)^{-2/3}$~MeV.
This result does not depend on the heavy neutrino mass because they
annihilate with massless active ones, $\nu_2 +\bar \nu_a
\rar all$. The heavy neutrinos would be relativistic at decoupling and
their number density would not be Boltzmann suppressed if, say, 
$T_f>M_s/2$. This gives  
\be 
\sin^2 2\theta (\dm/\mbox{MeV}^2)^{3/2} < 500~.
\label{relath}
\ee
If this condition is not fulfilled the impact of $\nu_s$ on BBN
would be strongly diminished. On the other hand, for a sufficiently
large mass and non-negligible mixing, the $\nu_2$ lifetime given by
Eq.~(\ref{taumuh}) would be quite short, so that they would all decay
prior to the BBN epoch.  (To be more exact, their number density would
not be frozen, but follow the equilibrium form 
$\propto e^{-m_2/T_\gamma}$.)

Another possible effect that could diminish the impact of heavy
neutrinos on BBN is entropy dilution. If $\nu_2$ were decoupled while
being relativistic, their number density would not be suppressed
relative to light active neutrinos. However, if the decoupling
temperature is higher than, say, 50~MeV pions and muons were still
abundant in the cosmic plasma and their subsequent annihilation would
diminish the relative number density of heavy neutrinos. If the
decoupling temperature is below the QCD phase transition the dilution
factor is at most $17.25/10.75 =1.6$.  Above the QCD phase transition
the number of degrees of freedom in the cosmic plasma is much larger
and the dilution factor is approximately 5.5. However, these effects
are essential for very weak mixing, for example the decoupling
temperature would exceed 200~MeV if $\sin^2 2\theta < 8\times
10^{-6}$. For such a small mixing the life-time of the heavy
$\nu_2$ would exceed the nucleosynthesis time and they would be
dangerous for BBN even if their number density is 5 times diluted.

Sterile neutrinos would never be abundant in the
universe if $\Gamma_s/H < 1$. In fact we can impose a stronger
condition demanding that the energy density of heavy neutrinos
should be smaller than the energy density of one light neutrino
species at BBN ($T\sim 1$ MeV). Taking into account a possible entropy
dilution by factor 5  we obtain the 
bound:
\be
\left( \dm/\mbox{MeV}^2\right)\,\sin^2 2\theta 
< 2.3\times 10^{-7} \, .
\label{dmsinh}
\ee
Parameters satisfying this conditions cannot be excluded by BBN.

If $\nus$ (or to be more precise $\nu_2$)
mass is larger than 135 MeV, the dominant decay mode becomes
$\nu_2 \rar \pi^0 + \nu_a$. The life-time with respect to this decay can
be found from the calculations~\cite{fishbach77,kalogeropoulous79}
of the decay rate $\pi^0 \rar \nu \bar \nu $ and is equal to:
\be
\tau = \left[{ G_F^2 m_s (m_s^2 -m_\pi^2) f_\pi^2 \sin^2\theta 
\over 16\pi} \right]^{-1}
= 5.8 \cdot 10^{-9} \,{\rm sec} \left[ \sin^2 \theta { m_s (m_s^2-m_\pi^2)
\over m_\pi^3} \right]^{-1}
\label{taupi}
\ee
where $m_s\equiv m_2$ is the mass of the sterile neutrino (we always
assume that the mixing angle is small so that $\nu_2\approx \nus$), 
$m_\pi = 135$ MeV is the $\pi^0$-mass and $f_\pi= 131 $ MeV is
the coupling constant for the decay $\pi^+ \rar \mu+\nu_\mu$. 
The approximate estimates of ref.~\cite{dolgov00hv} permit one to conclude
that for the life-time of $\nu_2$ smaller than 0.1 sec, and corresponding
cosmological temperature higher than 3 MeV, the decay products would
quickly thermalize and their impact on BBN would be small. For a life-time
larger than 0.1 sec, and $T<3$ MeV, one may assume that thermalization of
neutrinos is negligible and approximately evaluate their impact on BBN.
If $\nu_s$ is mixed with $\num$ or $\nut$ then electronic neutrinos are
not produced in the decay $\nus\rar \pi^0 \nu_a$ and only 
the contribution of the
decay products into the total energy density is essential. As we have
already mentioned, non-equilibrium $\nue$ produced by the decay would
directly change the frozen $n/p$-ratio. This case is more complicated
and demands a more refined treatment.

The $\pi^0$ produced in the decay $\nus\rar \nu_a+\pi^0$ immediately decays  
into two photons and they heat up the electromagnetic part of the plasma, 
while neutrinos by assumption are decoupled from it. We estimate the
fraction of energy delivered into the electromagnetic and neutrino
components of the cosmic plasma in the instant decay approximation.
Let $r_s=n_s/n_0 $ be the ratio of the number densities of the heavy  
neutrinos with respect to the equilibrium light ones, 
$n_0 = 0.09 T_\gamma^3$. The total energy of
photons and $e^+e^-$-pairs including the photons produced by the
decay is
\be
\rho_{em} = {11\over 2}{\pi^2 \over 30} T^4 + r_s n_0 {m_s \over 2} 
\left( 1+{m_\pi^2 \over m_s^2}\right)\, ,
\label{rhoemh}
\ee
while the energy density of neutrinos is
\be
\rho_{\nu} = {21\over 4}{\pi^2 \over 30} T^4 + r_s n_0 {m_s \over 2} 
\left( 1- {m_\pi^2 \over m_s^2}\right) \,.
\label{rhonuh}
\ee
The effective number of neutrino species at BBN can be defined as
\be
N_\nu^{(eff)} = {22\over 7} {\rho_\nu \over \rho_{em}} \, .
\label{keffnuh}
\ee
Because of the stronger heating of the electromagnetic component of the
plasma by the decay products, the relative role of neutrinos diminishes
and $N_\nu^{(eff)}$ becomes considerably smaller than 3. If $\nu_s$ are
decoupled while relativistic their fractional number 
at the moment of decoupling would be
$r_s=4$ (two spin states and antiparticles are included). Possible
entropy dilution could diminish it to slightly below 1. Assuming that the 
decoupling temperature of weak integrations is $T_W = 3$ MeV we find that 
$N_\nu^{(eff)} =0.6 $ for $m_s = 150$ MeV and 
$N_\nu^{(eff)} =1.3 $ for $m_s = 200$ MeV, if the frozen number density
of $\nu_s$ is not diluted by the later entropy release and $r_s$ remains
equal to 4. If it was diluted down to 1, then the numbers would 
respectively change to $N_\nu^{(eff)} =1.15 $ for $m_s = 150$ MeV and 
$N_\nu^{(eff)} =1.7 $ for $m_s = 200$ MeV, instead of the standard
$N_\nu^{(eff)} =3 $. 
Thus a very heavy $\nu_s$ would result in under-production of $^4 He$.
There could, however, be some other effects acting in the opposite direction.

Since $\nue$ decouples from electrons/positrons at smaller temperature
than $\num$ and $\nut$, the former may have enough time to thermalize.
In this case the temperatures of $\nue$ and photons would be the same
(before $e^+e^-$-annihilation) and the results obtained above would be
directly applicable. However, 
if thermalization between $\nue$ and $e^\pm$ was not efficient, then 
the temperature of electronic neutrinos at BBN would be smaller than in
the standard model. The deficit of $\nue$ would produce an opposite
effect, namely enlarging the production of primordial $^4 He$, because it
results in an increase of the $n/p$-freezing temperature. This effect
significantly dominates the decrease of $N_\nu^{(eff)}$ discussed above.
Moreover even in the case of the decay 
$\nu_2 \rar \pi^0 + \nu_{\mu, \tau}$, when $\nue$ are not directly 
created through the decay,
the spectrum of the latter may be distorted at the high energy tail
by the interactions with
non-equilibrium $\nut$ and $\num$ produced by the decay. This would
result in a further increase of $^4He$-production. In the case of 
direct production of non-equilibrium $\nue$ through the decay
$\nu_2 \rar \pi^0 + \nu_e$ their impact on $n/p$ ratio would be even 
much stronger. 

To summarize, there are several different effects from the decay of $\nu_s$
into $\pi^0$ and $\nu$ on BBN. Depending upon the decay life-time
and the channel these effects may operate in opposite directions. 
If the life-time of $\nu_2$ is  larger  than 0.1 sec but smaller than
0.2 sec, so that $e^\pm$ and $\nue$ establish equilibrium the production
of $^4He$ is considerably diminished so that this life-time interval
would be mostly forbidden. For life-times larger than 0.2 sec the
dominant effect is a decrease of the energy density of $\nue$ and this
results in a strong increase of the mass fraction of $^4 He$. Thus large 
life-times should also be forbidden. Of course there is a 
small part of the parameter space where both effects cancel each other
and this interval of mass/mixing is allowed. It is, however, difficult to
establish its precise position with the approximate arguments used in
ref.~\cite{dolgov00hv}.

Thus, in the case of $\nu_s \leftrightarrow \nu_{\mu, \tau}$ mixing for 
$m_s>140$ MeV we can exclude the life-times of 
$\nu_s$ roughly larger than 0.1~sec, except for a small region near
0.2~sec where two opposite effects cancel and 
the BBN results remain undisturbed despite the presence of sterile neutrinos.
Translating these results into mixing angle according to
eq. (\ref{taupi}), we conclude that mixing angles 
$\sin^2 \theta < 5.8 \cdot 10^{-8} m_\pi/m_s /((m_s/m_\pi)^2-1)$
are excluded by BBN. Combining this result with eq. (\ref{dmsinh})
we obtain the exclusion region for  $m_s>140 MeV$:
\be
5.1 \cdot 10^{-8} ~\frac{\mbox{MeV}^2}{m_s^2} 
<\sin^2 \theta < 5.8 \cdot 10^{-8}
~\frac{m_\pi}{m_s}\frac{1}{(m_s/m_\pi)^2-1}~.
\label{sin_pi_excl}
\ee 
   
In the case of $\nu_s \leftrightarrow \nu_e$ mixing 
for $m_s>140 MeV$ the limits are possibly stronger,
but it is more difficult to obtain reliable estimates because 
of a strong influence of non-equilibrium $\nu_e$, produced by the
decay, on neutron-proton reactions.

The constraints on the mass/mixing of $\nus$ from neutrino observation
of SN 1987A are analyzed in some detail
in ref.~\cite{dolgov00krm} and based on the upper limit of the 
energy loss into a new invisible channel because the latter would 
shorten the neutrino burst from this supernova below the observed 
duration.

\begin{figure}[t]
 \vspace{9.0cm}
\includegraphics{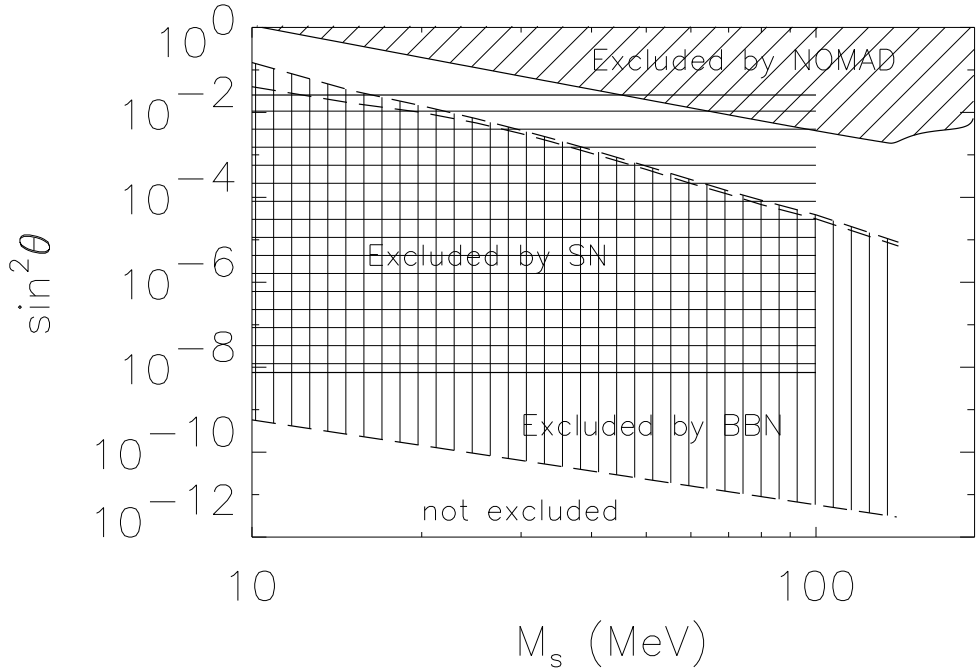}
 \caption{\it
Summary of the exclusion regions in the 
$(\sin^2 \theta$-$M_s)$-plane.
    \label{fig:all} }
\end{figure}

The results are summarized in fig.~\ref{fig:all}. 
The region between the two horizontal lines
running up to 100 MeV are excluded by the duration of the neutrino
burst from SN~1987A.  BBN excludes the
area below the two upper dashed lines if the heavy neutrinos were abundant
in the early universe. These two upper dashed lines both correspond to 
the conservative limit of one extra light neutrino species permitted
by the primordial $^4$He-abundance. The higher of the two is for  
mixing with $\nu_{\mu, \tau}$ and the slightly lower curve is for 
mixing with $\nu_e$.
In the region below the lowest dashed curve the
heavy neutrinos are not efficiently produced in the early universe and
their impact on BBN is weak. 
For comparison we have also presented here the region excluded by NOMAD
Collaboration~\cite{nomad00} for the case of
$\nu_s \leftrightarrow \nu_\tau$ mixing.
A more accurate consideration would probably permit
to expand the excluded region both in the horizontal and vertical
directions.

\section{Acknowledgements}
I am greateful to P. Di Bari for the discussion of the formation of
leptonic domains.

%
\end{document}